\documentclass[10pt,twoside,a4]{article}
\usepackage{amssymb}
\usepackage{amsfonts}
\usepackage{amsmath}
\usepackage{graphicx}
\usepackage{url}
\usepackage{xcolor}

\newlength{\figurewidth}
\newlength{\figureheight}
\definecolor{amundi_blue}{RGB}{0,176,240}
\definecolor{amundi_dark_blue}{RGB}{0,28,75}

\newif\ifResearchVersion
\ResearchVersiontrue

\begin{document}

\ifResearchVersion

\setcounter{page}{1}

\title{\textbf{\color{amundi_blue}The Market Measure of Carbon Risk and\\its Impact on the Minimum Variance Portfolio}%
\footnote{The authors are grateful to Martin Nerlinger from the University of
Augsburg, who provided them with the time series of the BMG risk
factor (see \texttt{https://carima-project.de/en/downloads} for more
details about this carbon risk factor). They would also like to
thank Melchior Dechelette, Lauren Stagnol and Bruno Taillardat for
their helpful comments.}}
\author{
{\color{amundi_dark_blue} Th\'eo Roncalli} \\
Master in Bioinformatics \& Biostatistics \\
University Paris-Saclay, Paris \\
\texttt{theo.roncalli@universite-paris-saclay.fr} \and
{\color{amundi_dark_blue} Th\'eo Le Guenedal, Fr\'ed\'eric Lepetit, Thierry Roncalli, Takaya Sekine} \\
Quantitative Research \\
Amundi Asset Management, France \\
\texttt{firstname.lastname@amundi.com}}

\date{\color{amundi_dark_blue}January 2021}

\maketitle

\vspace*{-10pt}
\begin{abstract}
Like ESG investing, climate change is an important concern for asset
managers and owners, and a new challenge for portfolio construction.
Until now, investors have mainly measured carbon risk using
fundamental approaches, such as with carbon intensity metrics.
Nevertheless, it has not been proven that asset prices are directly
impacted by these fundamental-based measures. In this paper, we
focus on another approach, which consists in measuring the
sensitivity of stock prices with respect to a carbon risk factor. In
our opinion, carbon betas are market-based measures
that are complementary to carbon intensities or fundamental-based
measures when managing investment portfolios, because carbon betas
may be viewed as an extension or forward-looking measure of the
current carbon footprint. In particular, we show how this new
metric can be used to build minimum variance strategies and how they
impact their portfolio construction.
\end{abstract}

\medskip

\medskip

\noindent \textbf{Key findings:}
\begin{enumerate}
\item Measuring carbon risk is different if we consider a fundamental-based approach by using carbon intensity metrics or a market-based approach by using carbon betas.
\item Managing relative carbon risk implies to overweight green firms, whereas managing absolute carbon risk implies having zero exposure to the carbon risk factor. The first approach is an active management bet, while the second case is an immunization investment strategy.
\item Both specific and systematic carbon risks are important when building a minimum variance portfolio and justify combining fundamental and market approaches of carbon risk.
\end{enumerate}

\clearpage

\else

\setcounter{page}{7}

\fi

\section{INTRODUCTION}

According to Mark Carney (2019), climate change is one of the big
current challenges faced by the financial sector with the goal to
accelerate the transition to a low carbon economy. It concerns all
the financial institutions: central banks, commercial banks,
insurance companies, asset managers, asset owners, etc. Among the
several underlying topics, the risk management of climate change
will be one of the pillars of the future regulation in order to
ensure financial sector resilience to a tail risk. Since risk
management must concern both physical and transition risks (Carney,
2015), incorporating climate change when managing banks' credit
portfolios is not obvious. The question is how climate change
impacts the default probability of issuers. The same issue occurs
when we consider stock and bond portfolios of asset managers and
owners. Indeed, we have to understand how asset prices react to
climate change. This is why we have to develop new risk metrics in
order to assess the relationship between climate change and asset
returns. However, we face data collection issues when we consider
this broad subject. Therefore, we focus here on carbon risk since it
is the main contributor to climate change\footnote{This implies that
we consider  transition risks, not physical rirks.} and we have more
comprehensive and robust data on carbon metrics at the issuer level.%
\smallskip

The general approach to managing an investment portfolio's carbon
risk is to reduce or control the portfolio's carbon footprint, for
instance by considering CO2 or CO2e emissions. This approach assumes
that the carbon risk will materialize and that having a portfolio
with a lower exposure to CO2 emissions will help to avoid some
future losses. The main assumption of this approach is then to
postulate that firms that currently have high carbon footprints will
be penalized in the future in comparison with firms that currently
have low carbon footprints. In this paper, we use an alternative
approach. We define carbon risk from a financial point of view, and
we consider that the carbon risk of equities corresponds to the
market risk priced in by the stock market. This carbon financial
risk can be decomposed into a common (or systematic) risk factor and
a specific (or idiosyncratic) risk factor. Since identifying the
specific risk is impossible, we focus on the common risk factor that
drives the carbon risk. The objective is then to build a
market-based risk measure to manage the carbon risk in investment
portfolios. This is exactly the framework proposed by G\"orgen et
al. (2019) in their seminal paper.\smallskip

In this framework, the carbon financial risk of a stock corresponds to its
price sensitivity to the carbon risk factor. This carbon beta is a
market-based relative risk and may be viewed as an extension or
forward-looking measure of the carbon footprint, where the objective is to
be more exposed to green firms than to brown ones. In this case, this is
equivalent to promoting stocks with a negative carbon beta over stocks with
a positive carbon beta. This approach of relative carbon risk differs from
the approach of absolute carbon risk, which is measured at the stock level
by the absolute value of the carbon beta, because absolute carbon risk
considers that both large positive and negative carbon beta values incur a
financial risk that must be reduced. This is an agnostic or neutral method,
contrary to the first method which is more related to investors' moral
values or convictions.\smallskip

Since the 2008 Global Financial Crisis, institutional investors have
widely used minimum variance (MV) strategies to reduce their equity
investments' market risk. While the original idea of these
strategies was to reduce the portfolio's volatility, today the goal
of minimum variance strategies is to manage the largest financial
unrewarded risks and not just volatility risk. This is why
sophisticated MV programs also include idiosyncratic valuation risk,
reputational risk, etc. In this context, incorporating climate risk
into minimum variance portfolios is natural. Therefore, we propose a
two-factor model that is particularly adapted to this investment
strategy and show that the solution depends on whether we would like
to manage relative or absolute carbon risk.

\section{THE MARKET MEASURE OF CARBON RISK}

To manage a portfolio's carbon risk, carbon risk needs to be
measured at the company level. There are different ways to measure
this risk, including the fundamental and market approaches. In this
paper, we favor the second approach because it provides a better
assessment of the impact of climate-related transition risks on each
company's stock price. Moreover, the market-based approach allows us
to mitigate the issue of a lack of climate change-relevant
information. In what follows, we present this latest approach by
using the mimicking portfolio for carbon risk developed by G\"orgen
et al. (2019). We compare this seminal approach with a simplified
approach, which consists in using direct metrics such as carbon
intensity. Once carbon betas are computed, we can analyze the carbon
risk of each company priced in by the stock market and compare it
with the carbon intensity, which is the most used fundamental-based
measure of carbon risk. We also discuss the difference between
relative and absolute carbon risk.

\subsection{Measuring carbon risk}

Measuring a company's carbon risk using the carbon beta of its stock
price was first proposed by G\"orgen et al. (2019). In what follows,
we summarize their approach and test alternative approaches.
Moreover, we suggest using the Kalman filter in order to estimate
the dynamic carbon beta of stock prices.

\paragraph{The Carima approach}

The goal of the carbon risk management (Carima) project, developed
by G\"orgen et al. (2019), is to develop \textquotedblleft \textsl{a
quantitative tool in order to assess the opportunities of profits
and the risks of losses that occur from the transition
process}\textquotedblright. The Carima approach combines a
market-based approach and a fundamental approach. Indeed, the carbon
risk of a firm or a portfolio is measured by considering the
dynamics of stock prices which are partly determined by climate
policies and transition processes towards a green economy.
Nevertheless, a prior fundamental approach is important to quantify
carbon risk. In practical terms, the fundamental approach consists
in defining a carbon risk score for each stock in an investment
universe using a set of objective measures, whereas the market
approach consists in building a brown minus green or BMG carbon risk
factor, and computing the risk sensitivity of stock prices with
respect to this BMG factor. Therefore, the carbon factor is derived
from climate change-relevant information from numerous firms.%
\smallskip

In the Carima approach, the BMG factor is developed using a large
amount of climate-relevant information provided by different
databases. In the following, we detail the methodology used by the
Carima project to construct the BMG factor. Two steps are required
to develop this new common risk factor: (1) the development of a
scoring system to determine if a firm is green, neutral or brown and
(2) the construction of a mimicking factor portfolio for carbon risk
which has a long exposure to brown firms and a short exposure to
green firms. The first step consists in defining a brown green score
(BGS) using a fundamental approach to assess the carbon risk of
different firms. This scoring system uses four ESG databases over
the period from 2010 to 2016: Thomson Reuters ESG, MSCI ESG Ratings,
Sustainalytics ESG ratings and the Carbon Disclosure Project (CDP)
climate change questionnaire. Overall, $55$ carbon risk proxy
variables are retained. Each variable is transformed into a dummy
derived with respect to the median, meaning that $1$ corresponds to
a brown value and $0$ corresponds to a green value. Then, G\"orgen
et al. (2019) classified the variables into three different
dimensions that may affect the stock value of a firm in the event of
unexpected shifts towards a low carbon economy: (1) value chain, (2)
public perception and (3) adaptability. The value chain dimension
mainly deals with current emissions while the adaptability dimension
reflects potential future emissions determined in particular by
emission reduction targets and environmental R\&D spending. Then,
three scores are created and correspond to the average of all
variables contained in each dimension: the \textit{value chain}
$\mathrm{VC}$, the \textit{public perception} $\mathrm{PP}$ and the
\textit{non-adaptability} $\mathrm{NA}$. It follows that each score
has a range between $0$ and $1$. G\"orgen et al. (2019) proposed
defining the brown green score (BGS) using the following equation:
\begin{equation}
\mathrm{BGS}_{i}\left( t\right) =\frac{2}{3}\left( 0.7\cdot \mathrm{VC}_{i}\left( t\right) +
0.3\cdot \mathrm{PP}_{i}\left( t\right) \right) +
\frac{\mathrm{NA}_{i}\left( t\right) }{3}\left( 0.7\cdot \mathrm{VC}_{i}\left( t\right) +
0.3\cdot \mathrm{PP}_{i}\left( t\right)\right)
\label{eq:BGS}
\end{equation}
The higher the BGS value, the browner the firm. The second step
consists in constructing a brown minus green (BMG) risk factor. Here
the Carima project considers an average BGS for each stock that
corresponds to the mean value of the BGS over the period in
question, from 2010 to 2016. The construction of the BMG factor
follows the methodology of Fama and French (1992), which consists in
splitting the stocks into six portfolios: small green (SG), small
neutral (SN), small brown (SB), big green (BG), big neutral (BN) and
big brown (BB). Then, the return of the BMG factor is defined as
follows:
\begin{equation}
R_{\mathrm{bmg}}\left( t\right) =\frac{1}{2}\left( R_{\mathrm{SB}}\left( t\right) +R_{\mathrm{BB}}\left( t\right) \right) -
\frac{1}{2}\left( R_{\mathrm{SG}}\left( t\right) +R_{\mathrm{BG}}\left( t\right) \right)
\label{eq:BMG}
\end{equation}
where the returns of each portfolio are value-weighted.

\paragraph{Alternative approaches}

Since the Carima approach is based on $55$ variables from $4$ ESG
databases, it may be complicated for investors and academics to
reproduce the BMG factor of G\"orgen et al. (2019). This is why
Roncalli et al. (2020) proposed several proxies that may be easily
computed. They used the same approach to build the BMG factor, but
replaced the brown green score by simple scoring systems using a
single variable. Among the different tested factors\footnote{To
build these factors, they have considered the stocks that were
present in the MSCI World index during the 2010-2018 period.},
Roncalli et al. (2020) showed that the Carima BMG factor is highly
correlated to two BMG factors based on (1) the carbon intensity
derived on the three scopes (Trucost dataset) and (2) the MSCI
carbon emissions exposure score (MSCI, 2020). In Exhibit
\ref{fig:jpm_carbon1}, we report the cumulative performance of these
two factors and the Carima factor. We observe that the three factors
are very similar and highly correlated. On average, we observe that
brown firms slightly outperformed green firms from 2010 to 2012.
Then, the cumulative return fell by almost $35\%$ because of the
unexpected path in the transition process towards a low carbon
economy. From 2016 to the end of the study period, brown firms
created a slight excess performance. Overall, the best-in-class
green stocks outperform the worst-in-class green stocks over the
study period with an annual return of $2.52\%$ for the Carima
factor, $3.09\%$ for the carbon intensity factor and $4.01\%$ for
the factor built with the carbon emissions exposure score.

\begin{table}[tbh]
\centering
\caption{Cumulative performance of the BMG factors}
\label{fig:jpm_carbon1}
\smallskip

\includegraphics[width = \figurewidth, height = \figureheight]{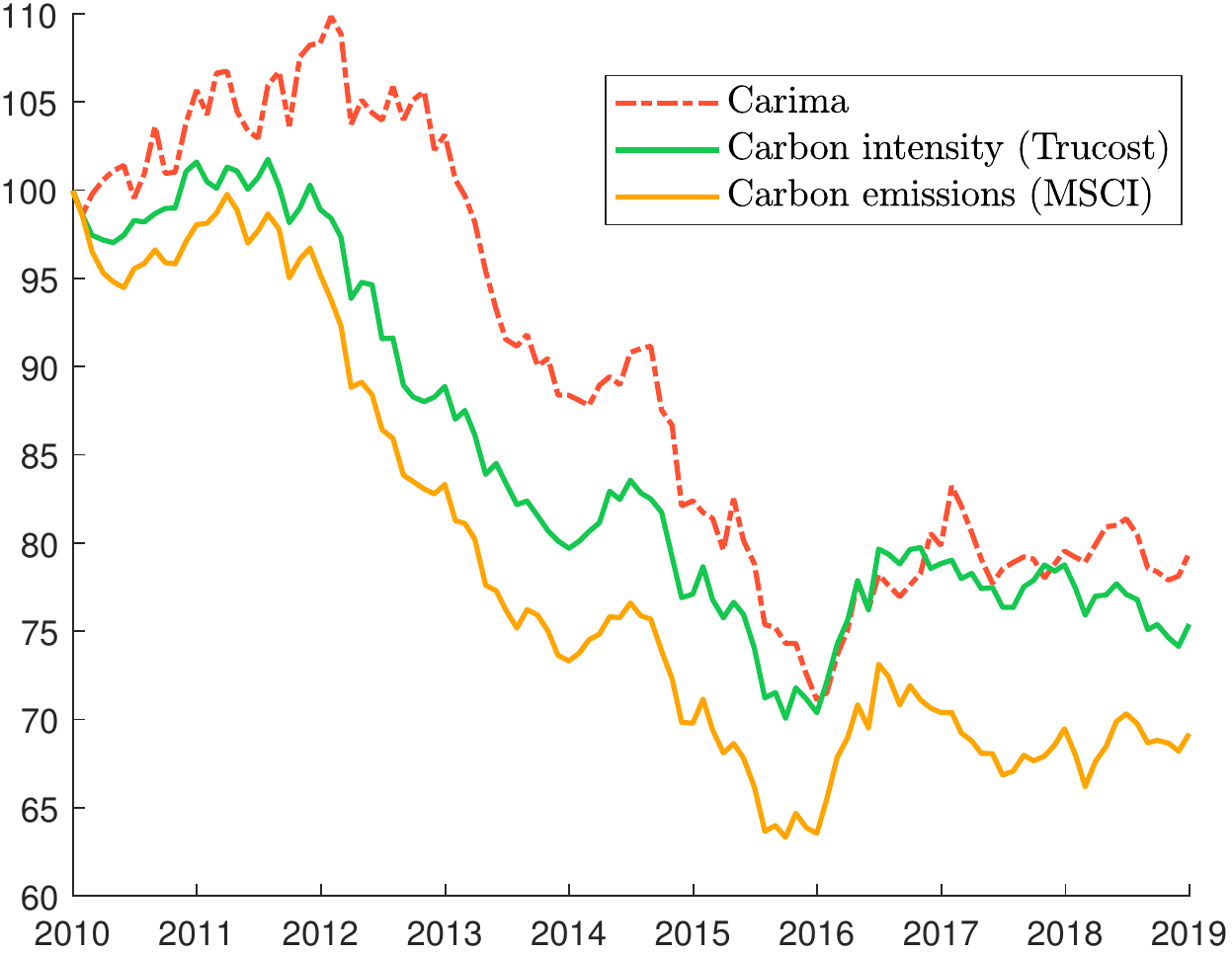}
\end{table}

\paragraph{Estimation of the carbon beta}

G\"orgen et al. (2019) and Roncalli et al. (2020) tested several
models to estimate the carbon beta by considering different sets of
risk factors that include market, size, value and momentum risk
factors. While the first authors used a static approach by assuming
that the carbon beta is constant over the period, the second authors
proposed a dynamic approach by assuming that the betas are
time-varying. This is more realistic since carbon betas may evolve
with the introduction of a climate-related policy, a firm's
environmental controversies, a change in the firm's environmental
strategy, an increased incorporation of carbon risk into portfolio
strategies, etc. In what
follows, we consider the dynamic approach with a two-factor model. Let $%
R_{i}\left( t\right) $ be the monthly return of stock $i$ at time $t$. We
assume that:%
\begin{equation}
R_{i}\left( t\right) =\alpha _{i}\left( t\right) +\beta _{\mathrm{mkt}%
,i}\left( t\right) R_{\mathrm{mkt}}\left( t\right) +\beta _{\mathrm{bmg}%
,i}\left( t\right) R_{\mathrm{bmg}}\left( t\right) +\varepsilon _{i}\left(
t\right)
\end{equation}%
where $R_{\mathrm{mkt}}\left( t\right) $ is the return of the market risk
factor, $R_{\mathrm{bmg}}\left( t\right) $ is the return of the BMG factor
and $\varepsilon _{i}\left( t\right) \sim \mathcal{N}\left( 0,\tilde{\sigma}
_{i}^{2}\right) $ is a white noise. The alpha component and the beta
sensitivities follow a random walk:%
\begin{equation}
\left\{
\begin{array}{l}
\alpha _{i}\left( t\right) =\alpha _{i}\left( t-1\right) +\eta _{\mathrm{%
alpha},i}\left( t\right)  \\
\beta _{\mathrm{mkt},i}\left( t\right) =\beta _{\mathrm{mkt},i}\left(
t-1\right) +\eta _{\mathrm{mkt},i}\left( t\right)  \\
\beta _{\mathrm{bmg},i}\left( t\right) =\beta _{\mathrm{bmg},i}\left(
t-1\right) +\eta _{\mathrm{bmg},i}\left( t\right)
\end{array}%
\right.
\end{equation}%
where $\eta _{\mathrm{alpha},i}\left( t\right) \sim
\mathcal{N}\left( 0,\sigma _{\mathrm{alpha},i}^{2}\right) $, $\eta
_{\mathrm{mkt},i}\left( t\right) \sim \mathcal{N}\left( 0,\sigma
_{\mathrm{mkt},i}^{2}\right) $ and $\eta _{\mathrm{bmg},i}\left(
t\right) \sim \mathcal{N}\left( 0,\sigma
_{\mathrm{bmg},i}^{2}\right) $ are three independent white noise
processes.\smallskip

In the sequel of the paper, we use the Carima factor to estimate the
carbon beta. For the market factor, we use the time series provided
by Kenneth French on his website. We estimate $\alpha _{i}\left(
t\right) $, $\beta _{\mathrm{mkt},i}\left( t\right) $ and $\beta
_{\mathrm{bmg},i}\left( t\right) $ for the stocks that belong to the
MSCI World index between January 2010 and December
2018\footnote{More precisely, we only consider the stocks that were
in the MSCI World index for at least three years during the
2010-2018 period and we take into account only the returns for the
period during which the stock is in the MSCI World index.} using the
Kalman filter (Fabozzi and Francis, 1978). Moreover, we scale the
Carima risk factor so that it has the same volatility as the market
risk factor over the entire period, implying that the magnitude of
the carbon beta $\beta _{\mathrm{bmg},i}\left( t\right) $ may be
understandable and comparable to the magnitude of the market beta
$\beta _{\mathrm{mkt},i}\left( t\right) $.\smallskip

The average carbon beta of a stock is equal to $0.05$, which is
close to zero, whereas the monthly variation of the carbon beta has
a standard deviation of $6.24\%$. If we consider the market beta,
the figures become respectively $1.02$ and $5.45\%$. Therefore, the
time volatility of the carbon beta is larger than the time
volatility of the market beta. In Exhibit \ref{fig:jpm_carbon2}, we
report the GICS sector analysis of the carbon sensitivities at the
end of December 2018. The box plots provide the median, the
quartiles and the $5\%$ and $95\%$ quantiles of the carbon beta. We
notice that on average, the energy, materials and real estate sector
have a positive carbon beta\footnote{This is in line with Bouchet
and Le Guenedal (2020) who demonstrated that credit risks are more
material in the energy and materials sectors. Therefore, the market
perceives these sectors as the entry points for systemic financial
carbon risks.} whereas the other sectors have a neutral or negative
carbon beta. The results differ slightly from those obtained by
G\"orgen et al. (2019) and Roncalli et al. (2020), who provided a
sector analysis by considering a constant carbon beta over the
period 2010--2018.\smallskip

\begin{table}[h]
\centering
\caption{Box plots of the dynamic carbon betas at the end of 2018}
\label{fig:jpm_carbon2}
\smallskip

\includegraphics[width = \figurewidth, height = \figureheight]{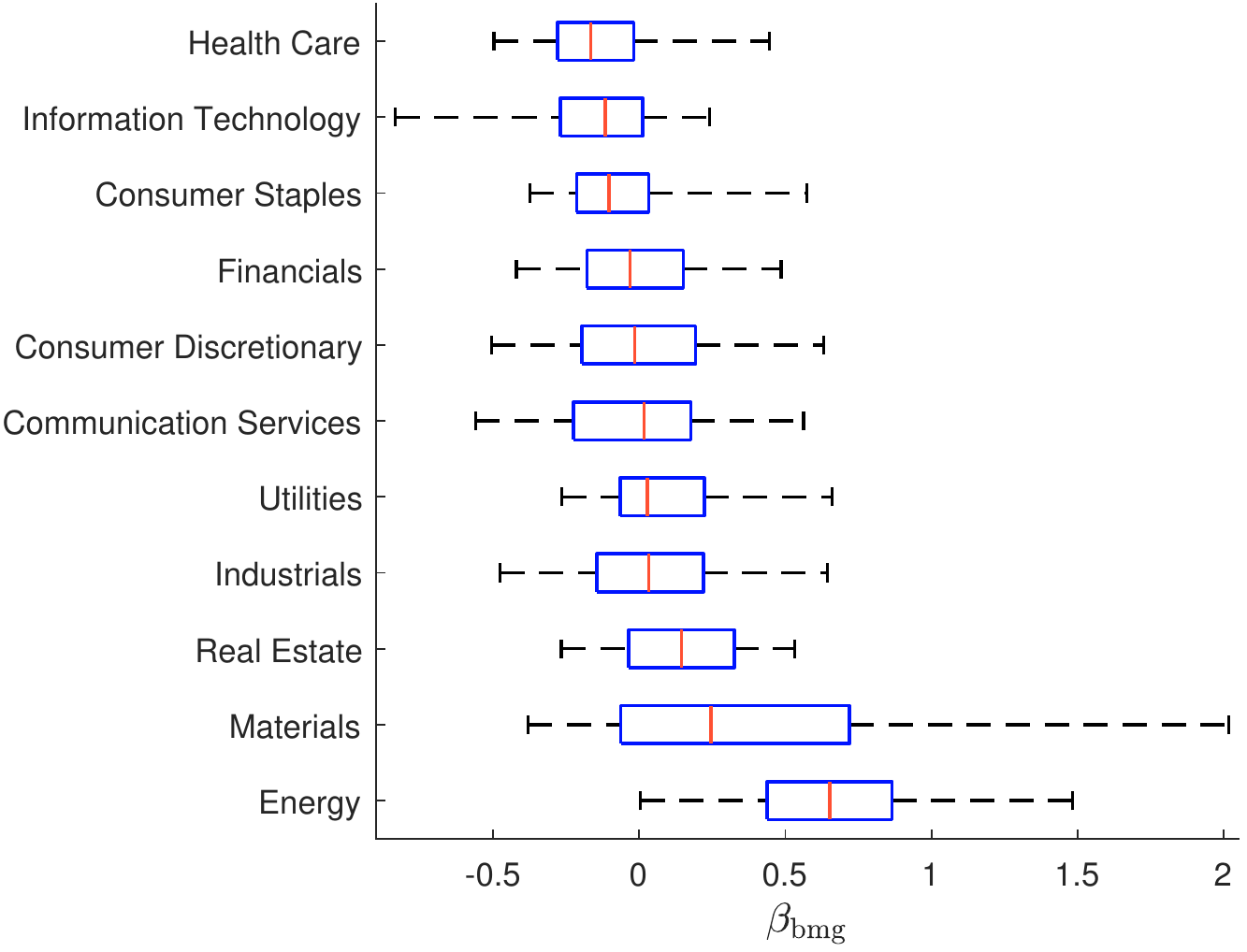}
\end{table}

The average carbon beta $\beta _{\mathrm{bmg},\mathcal{R}}\left( t\right) $ for the
region $\mathcal{R}$ at time $t$ is calculated as follows:
\begin{equation*}
\beta _{\mathrm{bmg},\mathcal{R}}\left( t\right) =\frac{\sum_{i\in \mathcal{R}}\beta
_{\mathrm{bmg},i}\left( t\right) }{\mathop{\rm card}\mathcal{R}}
\end{equation*}%
In Exhibit \ref{tab:jpm_carbon4a}, we report $\beta _{\mathrm{bmg},\mathcal{R}}\left(
t\right) $ for several MSCI universes at the end of each year: World (WD),
North America (NA), EMU, Europe-ex-EMU (EU) and Japan (JP). Whatever the
study period, the carbon beta $\beta _{\mathrm{bmg},\mathcal{R}}\left( t\right) $ is
positive in North America, which implies that American stocks are negatively
influenced by an acceleration in the transition process towards a green
economy. The average carbon beta is always negative in the Eurozone.
Overall, the Eurozone has always a lower average carbon beta than the world
as a whole, whereas the opposite is true for North America. Nevertheless,
the negative sensitivity of European equity returns has dramatically
decreased since 2010 and the BMG betas are getting closer for North America and the
Eurozone.

\begin{table}[h]
\centering
\caption{Relative carbon risk by region (end of year)}
\label{tab:jpm_carbon4a}
\smallskip

\begin{tabular}{crrrrr}
\hline
Year  & \multicolumn{1}{c}{WD} & \multicolumn{1}{c}{NA} &
\multicolumn{1}{c}{EMU} & \multicolumn{1}{c}{EU} & \multicolumn{1}{c}{JP} \\
\hline
2010  &   -0.02   &   0.13  &   -0.47  &   -0.16  &    0.02  \\
2011  &   -0.04   &   0.11  &   -0.48  &   -0.14  &   -0.07  \\
2012  &   -0.04   &   0.05  &   -0.32  &   -0.04  &   -0.13  \\
2013  &   -0.02   &   0.07  &   -0.21  &    0.05  &   -0.22  \\
2014  &   -0.04   &   0.01  &   -0.21  &   -0.02  &   -0.14  \\
2015  &    0.00   &   0.04  &   -0.20  &    0.05  &   -0.08  \\
2016  &    0.02   &   0.10  &   -0.21  &    0.01  &   -0.09  \\
2017  &    0.03   &   0.12  &   -0.23  &   -0.03  &   -0.04  \\
2018  &    0.06   &   0.10  &   -0.08  &    0.07  &   -0.02  \\
\hline
\end{tabular}
\end{table}

\subsection{Absolute versus relative carbon risk}

In the previous paragraph, the relative carbon risk of a stock $i$ at time $t$
is measured by its carbon beta value:
\begin{equation*}
\mathcal{RCR}_{i}\left( t\right) =\beta _{\mathrm{bmg},i}\left( t\right)
\end{equation*}
A majority of investors will prefer stocks with a negative carbon
beta over stocks with a positive carbon beta. However, an investment
portfolio with a negative carbon beta is exposed to the risk that
brown firms outperform green firms. In this case, reducing the
portfolio's  carbon risk means having a carbon beta as close as
possible to zero. This is why we introduce the concept of absolute
carbon risk, which is equal to the absolute value of the carbon
beta:
\begin{equation*}
\mathcal{ACR}_{i}\left( t\right) =\left\vert \beta _{\mathrm{bmg},i}\left(
t\right) \right\vert
\end{equation*}%
Exhibit \ref{fig:jpm_carbon3} presents the sector analysis of the absolute
carbon risk at the end of December 2018. From this point of view, utilities
is the least exposed sector to absolute carbon risk, whereas the energy and
materials are the sectors that are the most exposed.\smallskip

\begin{table}[h]
\centering
\caption{Box plots of the absolute carbon risk at the end of 2018}
\label{fig:jpm_carbon3}
\smallskip

\includegraphics[width = \figurewidth, height = \figureheight]{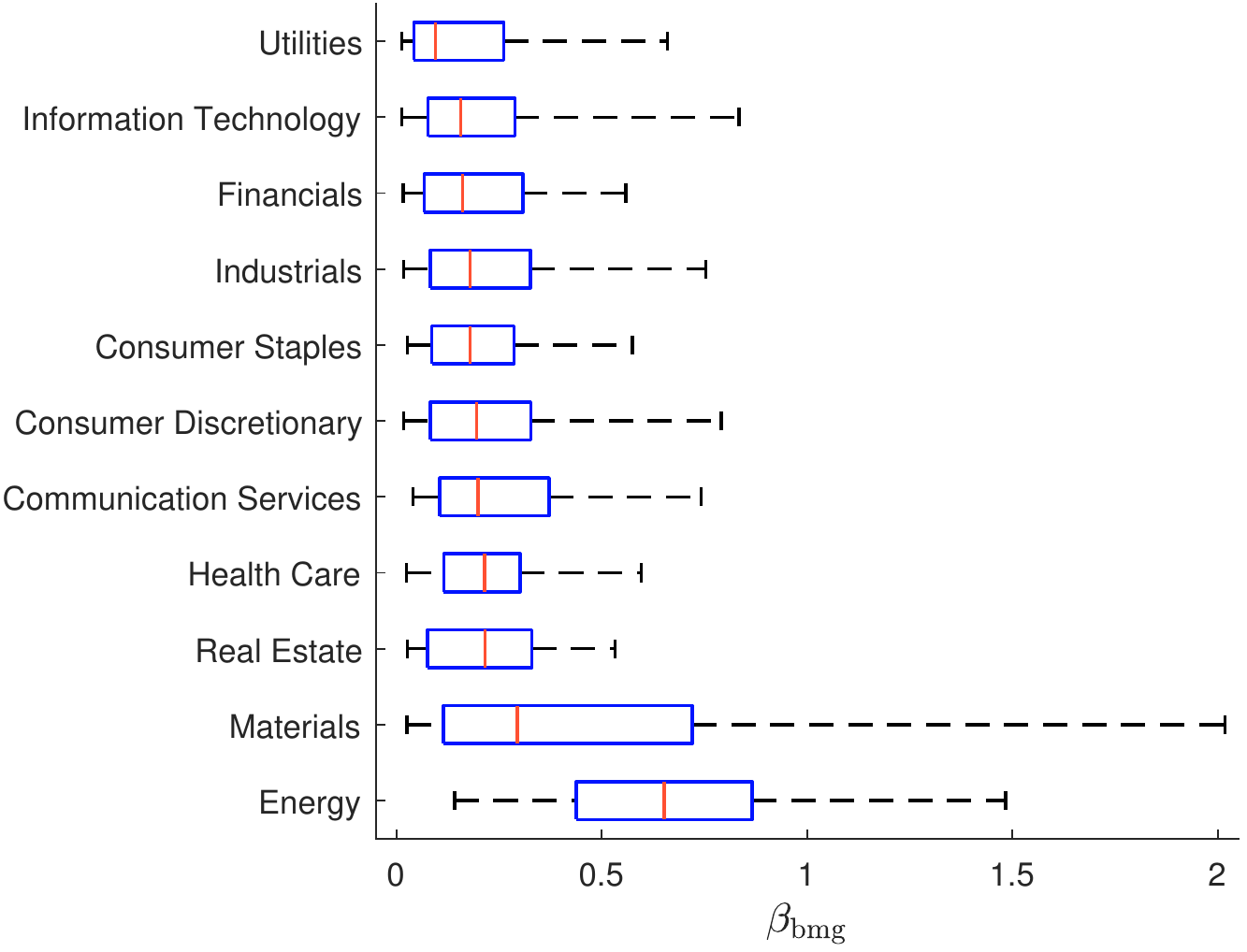}
\end{table}

$\mathcal{ACR}_{i}\left( t\right) $ is also a pricing magnitude measure of
the carbon risk. Let us consider an investment universe with two stocks. We
assume that $\beta _{\mathrm{bmg},1}\left( 1\right) =0.5$ and $\beta _{%
\mathrm{bmg},2}\left( 1\right) =-0.5$. On average, the relative
carbon risk is equal to zero, whereas the absolute carbon risk is
equal to $0.5$. One year later, we obtain $\beta
_{\mathrm{bmg},1}\left( 2\right) =1$ and $\beta
_{\mathrm{bmg},2}\left( 2\right) =-1$. In this case, the relative
carbon risk of the investment universe has not changed and is always
equal to zero. However, its absolute carbon risk has increased and
is now equal to $1$. It is obvious that the carbon risk is priced in
more in the second period than in the first period.\smallskip

We have reported the absolute carbon risk by region in Exhibit
\ref{tab:jpm_carbon4b}. We notice that the carbon risk was priced in
more in 2011 and 2012, because of the pricing magnitude in the
Eurozone. In this region, the absolute carbon risk has dramatically
decreased from $50\%$ in 2011 to $27\%$ in 2018. More globally, we
observe a convergence between the different developed regions. One
exception is Japan, where the absolute carbon risk is $50\%$ lower
than in Europe and North America.

\begin{table}[h]
\centering
\caption{Absolute carbon risk by region (end of year)}
\label{tab:jpm_carbon4b}
\smallskip

\begin{tabular}{cccccc}
\hline
Year &   WD &   NA &  EMU &   EU &   JP \\
\hline
2010 & 0.35 & 0.32 & 0.50 & 0.35 & 0.30 \\
2011 & 0.34 & 0.32 & 0.51 & 0.32 & 0.31 \\
2012 & 0.28 & 0.24 & 0.40 & 0.24 & 0.27 \\
2013 & 0.28 & 0.26 & 0.31 & 0.24 & 0.30 \\
2014 & 0.27 & 0.26 & 0.30 & 0.24 & 0.26 \\
2015 & 0.27 & 0.29 & 0.27 & 0.27 & 0.20 \\
2016 & 0.29 & 0.31 & 0.30 & 0.30 & 0.20 \\
2017 & 0.27 & 0.29 & 0.30 & 0.28 & 0.20 \\
2018 & 0.28 & 0.29 & 0.27 & 0.29 & 0.20 \\
\hline
\end{tabular}
\end{table}

\subsection{Comparison between market and fundamental measures of carbon
risk}

ESG rating agencies have developed many fundamental measures and
scores to assess a firm's carbon risk. For instance, the most
well-known is the carbon intensity $\mathcal{CI}_{i}\left(t\right)$,
which involves scopes 1, 2 and 3. In this paper, a firm's carbon
risk corresponds to the carbon beta priced in by the financial
market. It is not obvious that there is a strong relationship
between fundamental and market measures, because we may observe wide
discrepancies between the market perception of the carbon risk and
the carbon intensity of the firm. For instance, the linear
correlation between $\mathcal{CI}_{i}\left(t\right)$ and
$\beta_{\mathrm{bmg},i}\left(t\right)$ is equal to $17.4\%$ at the
end of December 2018. If we consider the BMG factor built directly
with the carbon intensity (Exhibit \ref{fig:jpm_carbon1}), the
correlation increases but remains relatively low since it is equal
to $21.4\%$ at the end of December 2018. The relationship between
$\mathcal{CI}_{i}\left(t\right)$ and
$\beta_{\mathrm{bmg},i}\left(t\right)$ is then more complex as seen
in Exhibit \ref{fig:jpm_carbon5}.\smallskip

\begin{table}[h]
\centering
\caption{Scatter plot of $\mathcal{CI}_{i}\left(t\right)$ and
$\beta_{\mathrm{bmg},i}\left(t\right)$ at the end of 2018}
\label{fig:jpm_carbon5}
\smallskip

\includegraphics[width = \figurewidth, height = \figureheight]{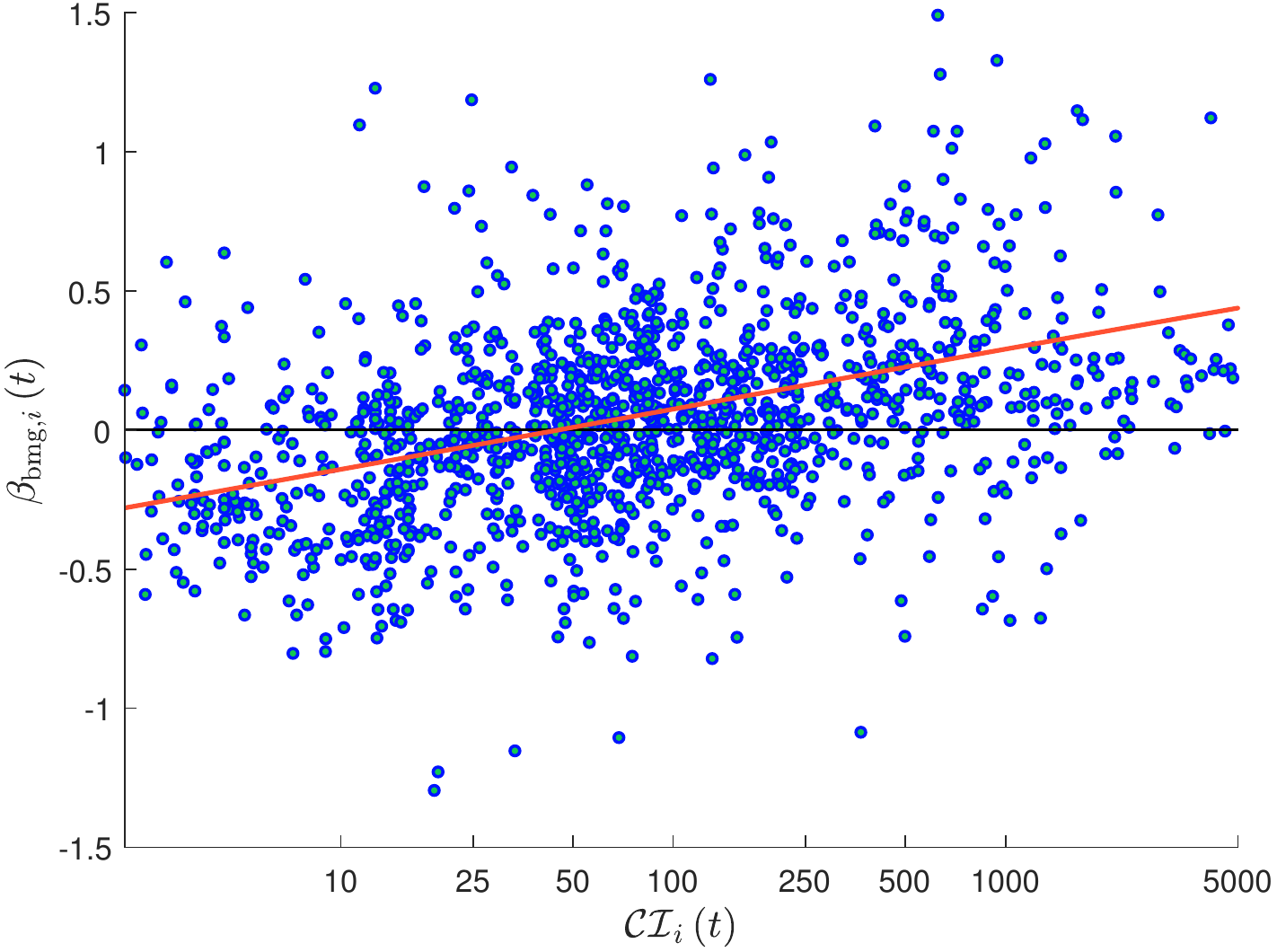}
\end{table}

This result is easily understandable because the stock market
incorporates dimensions other than carbon intensity to price in
the carbon risk. From a fundamental point of view, if the carbon
intensity of two firms is equal to $100$, they present the same
carbon risk. Nevertheless, we know that their risks depend on other
factors and parameters. For instance, it is difficult to compare two
firms with the same carbon intensity if they belong to two different
sectors or countries. The trajectory of the carbon intensity is also
another important factor. For instance, the risk is not the same if
one firm has dramatically decreased its carbon intensity in recent
years. Moreover, the adaptability issue, the capacity to transform
its business with investments in green R\&D or its financial
resources to absorb transition costs (Bouchet and Le Guenedal, 2020)
are other important parameters that impact the market perception of
the firm's carbon risk. Therefore, carbon intensity is less
appropriate to describe financial risks than carbon beta. In other
words, the carbon beta is an integrated measure of the different
fundamental factors affecting a firm's carbon risk.\smallskip

In Exhibit \ref{tab:jpm_carbon5}, we report the correlation between
$\mathcal{CI}_{i}\left(t\right)$ and
$\beta_{\mathrm{bmg},i}\left(t\right)$ at the end of December 2018.
We notice that it is higher in the Eurozone than in other regions.
In particular, the correlation in Japan is very low (less than
$5\%$). Moreover, we observe that it also differs with respect to
the sector. For instance, the financial sector presents the lowest
correlation value, certainly because the carbon risk of financial
institutions is less connected to their greenhouse gas emissions
than their (green and brown) investments and financing programs.

\begin{table}
\centering
\caption{Correlation in \% between $\mathcal{CI}_{i}\left(t\right)$ and
$\beta_{\mathrm{bmg},i}\left(t\right)$ at the end of 2018}
\label{tab:jpm_carbon5}
\smallskip

\begin{tabular}{lccccr}
\hline
Sector                      &   WD &   NA &  EMU &   EU & \multicolumn{1}{c}{JP} \\
\hline
Financials             & 18.2 & 20.1 & 29.2 & 20.9 & -1.5 \\
Energy                 & 18.2 & 17.8 & 31.8 & 24.5 &  3.8 \\
Materials              & 20.3 & 24.8 & 37.2 & 28.0 &  5.4 \\
Information Technology & 20.4 & 21.0 & 34.2 & 26.1 &  3.2 \\
Health Care            & 20.9 & 21.3 & 34.5 & 26.3 &  4.2 \\
Consumer Staples       & 21.5 & 22.3 & 35.1 & 26.4 &  4.3 \\
Communication Services & 21.6 & 22.2 & 32.2 & 24.7 &  6.6 \\
Consumer Discretionary & 22.3 & 23.1 & 37.8 & 25.8 &  2.6 \\
Real Estate            & 22.4 & 22.5 & 34.3 & 26.1 &  6.1 \\
Industrials            & 23.6 & 23.8 & 38.7 & 31.6 &  8.1 \\
Utilities              & 26.6 & 29.8 & 26.5 & 26.1 &  8.4 \\ \hline
All sectors            & 21.4 & 22.3 & 33.8 & 26.2 &  4.6 \\
\hline
\end{tabular}
\end{table}

\nopagebreak[4]

\section{INCORPORATING CARBON RISK INTO MINIMUM VARIANCE PORTFOLIOS}

There is an increasing appetite from fund managers of minimum
variance portfolios to take into account carbon risk because of two
main reasons. First, it is a financial and regulation risk that may
negatively impact stock returns. The second reason is that it is
highly sought after by institutional investors. In what follows, we
show how to incorporate carbon risk into these strategies. In
particular, we provide an analytical formula which is useful to
understand the impact of carbon betas on the minimum variance
portfolio and the covariance matrix of stock returns. We also
discuss the different practical implementations of minimum variance
portfolios when we consider market and fundamental measures of
carbon risk.

\subsection{Analytical results}

In this paragraph, we extend the famous formula of the minimum
variance portfolio when we complement the market risk factor with
the BMG factor. Then, we illustrate how the minimum variance
portfolio selects stocks in the presence of carbon risk.

\paragraph{Extension of the one-factor GMV formula}

We consider the global minimum variance (GMV) portfolio, which corresponds
to this optimization program:
\begin{eqnarray}
x^{\star } &=&\arg \min \frac{1}{2}x^{\top }\Sigma x  \label{eq:gmv1} \\
&\text{s.t.}&\mathbf{1}_{n}^{\top }x=1  \notag
\end{eqnarray}%
where $x$ is the vector of portfolio weights and $\Sigma $ is the covariance
matrix of stock returns. In the capital asset pricing model, we recall that:
\begin{equation}
R_{i}\left( t\right) =\alpha _{i}+\beta _{\mathrm{mkt},i}
R_{\mathrm{mkt}}\left( t\right) +\varepsilon _{i}\left( t\right)   \label{eq:gmv2}
\end{equation}%
where $R_{i}\left( t\right) $ is the return of asset $i$,
$R_{\mathrm{mkt}}\left( t\right) $ is the return of the market factor,
$\varepsilon _{i}\left( t\right) \sim \mathcal{N}\left(
0,\tilde{\sigma}_{i}^{2}\right) $ is the idiosyncratic risk and
$\tilde{\sigma}_{i}$ is the idiosyncratic volatility. Clarke et al. (2011) and
Scherer (2011) showed that:
\begin{equation}
x_{i}^{\star }=\frac{\sigma ^{2}\left( x^{\star }\right) }{\tilde{\sigma}%
_{i}^{2}}\left( 1-\frac{\beta _{\mathrm{mkt},i}}{\beta _{\mathrm{mkt}%
}^{\star }}\right)   \label{eq:gmv3}
\end{equation}%
where $\beta _{\mathrm{mkt}}^{\star }$ is a threshold and $\sigma
^{2}\left( x^{\star }\right) $ is the variance of the GMV portfolio.
Therefore, we note that the minimum variance portfolio is exposed to
stocks with low specific volatility $\tilde{\sigma}_{i}$ and low
beta $\beta _{\mathrm{mkt},i}$. More precisely, if asset $i$ has a
market beta $\beta _{\mathrm{mkt},i}$ smaller than the threshold
$\beta _{\mathrm{mkt}}^{\star }$, the weight of this asset is
positive: $x_{i}^{\star }>0$. If $\beta _{\mathrm{mkt},i}>\beta
_{\mathrm{mkt}}^{\star }$, then $x_{i}^{\star }<0$. Clarke et al.
(2011) extended Formula (\ref{eq:gmv3}) to the long-only case, where
$\beta _{\mathrm{mkt}}^{\star }$ is another threshold. In this case,
if $\beta_{\mathrm{mkt},i}<\beta _{\mathrm{mkt}}^{\star }$,
$x_{i}^{\star }>0$ and if $\beta _{\mathrm{mkt},i}\geq \beta
_{\mathrm{mkt}}^{\star }$, $x_{i}^{\star }=0$.\smallskip

We consider an extension of the CAPM by including the BMG risk factor:
\begin{equation}
R_{i}\left( t\right) =\alpha _{i}+\beta _{\mathrm{mkt},i}R_{\mathrm{mkt}}
\left( t\right) +\beta _{\mathrm{bmg},i}R_{\mathrm{bmg}}\left( t\right)
+\varepsilon _{i}\left( t\right)   \label{eq:gmv4}
\end{equation}%
where $R_{\mathrm{bmg}}\left( t\right) $ is the return of the BMG factor and
$\beta _{\mathrm{bmg},i}$ is the BMG sensitivity (or the carbon beta) of stock
$i$. Moreover, we assume that $R_{\mathrm{mkt}}\left( t\right) $ and
$R_{\mathrm{bmg}}\left( t\right) $ are uncorrelated. Roncalli et al. (2020)
showed that the GMV portfolio is defined as:
\begin{equation}
x_{i}^{\star }=\frac{\sigma ^{2}\left( x^{\star }\right) }{\tilde{\sigma}%
_{i}^{2}}\left( 1-\frac{\beta _{\mathrm{mkt},i}}{\beta _{\mathrm{mkt}}^{\star }}-\frac{\beta
_{\mathrm{bmg},i}}{\beta _{\mathrm{bmg}}^{\star }}\right)   \label{eq:gmv5}
\end{equation}%
where $\beta _{\mathrm{mkt}}^{\star }$ and $\beta _{\mathrm{bmg}}^{\star }$ are
two threshold values. In the case of long-only portfolios, we obtain a similar
formula:
\begin{equation}
x_{i}^{\star }=\frac{\sigma ^{2}\left( x^{\star }\right) }{\tilde{\sigma}%
_{i}^{2}}\max \left( 1-\frac{\beta _{\mathrm{mkt},i}}{\beta
_{\mathrm{mkt}}^{\star }}- \frac{\beta _{\mathrm{bmg},i}}{\beta
_{\mathrm{bmg}}^{\star }};0\right)   \label{eq:gmv6}
\end{equation}%
but with other values of the thresholds $\beta _{\mathrm{mkt}}^{\star }$ and $%
\beta _{\mathrm{bmg}}^{\star }$.

\paragraph{Interpretation of these results}

Contrary to the single-factor model, the impact of sensitivities is
more complex in the two-factor model. Indeed, we know that
$\bar{\beta}_{\mathrm{mkt}}\approx 1$ and
$\bar{\beta}_{\mathrm{bmg}}\approx 0$. It follows that $\beta
_{\mathrm{mkt}}^{\star }$ is positive, but $\beta
_{\mathrm{bmg}}^{\star }$ may be positive or
negative\footnote{Moreover, it generally takes a high absolute
value.}. We deduce that the ratio $\frac{\beta
_{\mathrm{mkt},i}}{\beta _{\mathrm{mkt}}^{\star }}$ is an increasing
function of $\beta _{\mathrm{mkt},i}$, but the ratio $\frac{\beta
_{\mathrm{bmg},i}}{\beta _{\mathrm{bmg}}^{\star }}$ may be an
increasing or a decreasing function of $\beta _{\mathrm{bmg},i}$.
The GMV portfolio will then always prefer stocks with low market
betas, but not necessarily stocks with low carbon betas. For
instance, it may prefer stocks with high carbon betas if $\beta
_{\mathrm{bmg}}^{\star }$ is negative.\smallskip

In the long-only case, a stock is selected if it satisfies the following
inequality:
\begin{equation*}
\frac{\beta _{\mathrm{mkt},i}}{\beta _{\mathrm{mkt}}^{\star }}+\frac{\beta _{\mathrm{bmg},i}}{\beta
_{\mathrm{bmg}}^{\star }}\leq 1
\end{equation*}%
Therefore, we notice that there is a trade-off between $\beta
_{\mathrm{mkt},i}$ and $\beta _{\mathrm{bmg},i}$. Nevertheless,
Roncalli et al. (2020) showed that the long-only MV portfolio tends
to prefer stocks with low absolute carbon risk.\smallskip

We recall that the volatility of stock $i$ is equal to $\sigma
_{i}^{2}=\beta _{\mathrm{mkt},i}^{2}\sigma _{\mathrm{mkt}}^{2}+\beta
_{\mathrm{bmg},i}^{2}\sigma
_{\mathrm{bmg}}^{2}+\tilde{\sigma}_{i}^{2}$, whereas the covariance
between stocks $i$ and $j$\ is equal to $\sigma _{i,j}^{2}=\beta
_{\mathrm{mkt},i}\beta _{\mathrm{mkt},j}\sigma
_{\mathrm{mkt}}^{2}+\beta _{\mathrm{bmg},i}\beta
_{\mathrm{bmg},j}\sigma _{\mathrm{bmg}}^{2}$. Therefore, choosing
stocks with low volatilities implies considering stocks with low
values of $\beta _{\mathrm{bmg},i}^{2}$. In a similar way, removing
stocks with high positive correlations implies removing stocks with
high values of $\beta _{\mathrm{bmg},i}\beta _{\mathrm{bmg},j}$.
This explains that the MV portfolio will prefer stocks with low
values of $\left\vert \beta _{\mathrm{bmg},i}\right\vert $.

\subsection{Practical implementations}

We now apply the previous framework to the MSCI World index at December 2018, and
illustrate the difference between absolute and relative carbon risk
when we consider the minimum variance portfolio.
Moreover, we compare these market-based approaches with implementations
of minimum variance portfolios that use fundamental carbon risk metrics.

\paragraph{Impact of carbon risk}

In Exhibit \ref{fig:jpm_carbon6}, we indicate the stocks that make up the MV
portfolio with respect to their beta values $\beta _{\mathrm{mkt},i}$ and
$\beta _{\mathrm{bmg},i}$. We find that the most important axis is the market
beta. Indeed, the market risk of a stock determines whether the stock is
included in the MV portfolio or not whereas the carbon risk adjusts the weights
of the asset. As we can see, the portfolio overweights assets whose market and
carbon sensitivities are both close to zero. This solution is satisfactory if
the original motivation is to reduce the portfolio's absolute carbon risk, but
it is not satisfactory if the objective is to manage the portfolio's relative
carbon risk.

\begin{table}[p]
\centering
\caption{Weights of the long-only MV portfolio}
\label{fig:jpm_carbon6}
\smallskip

\includegraphics[width = \figurewidth, height = \figureheight]{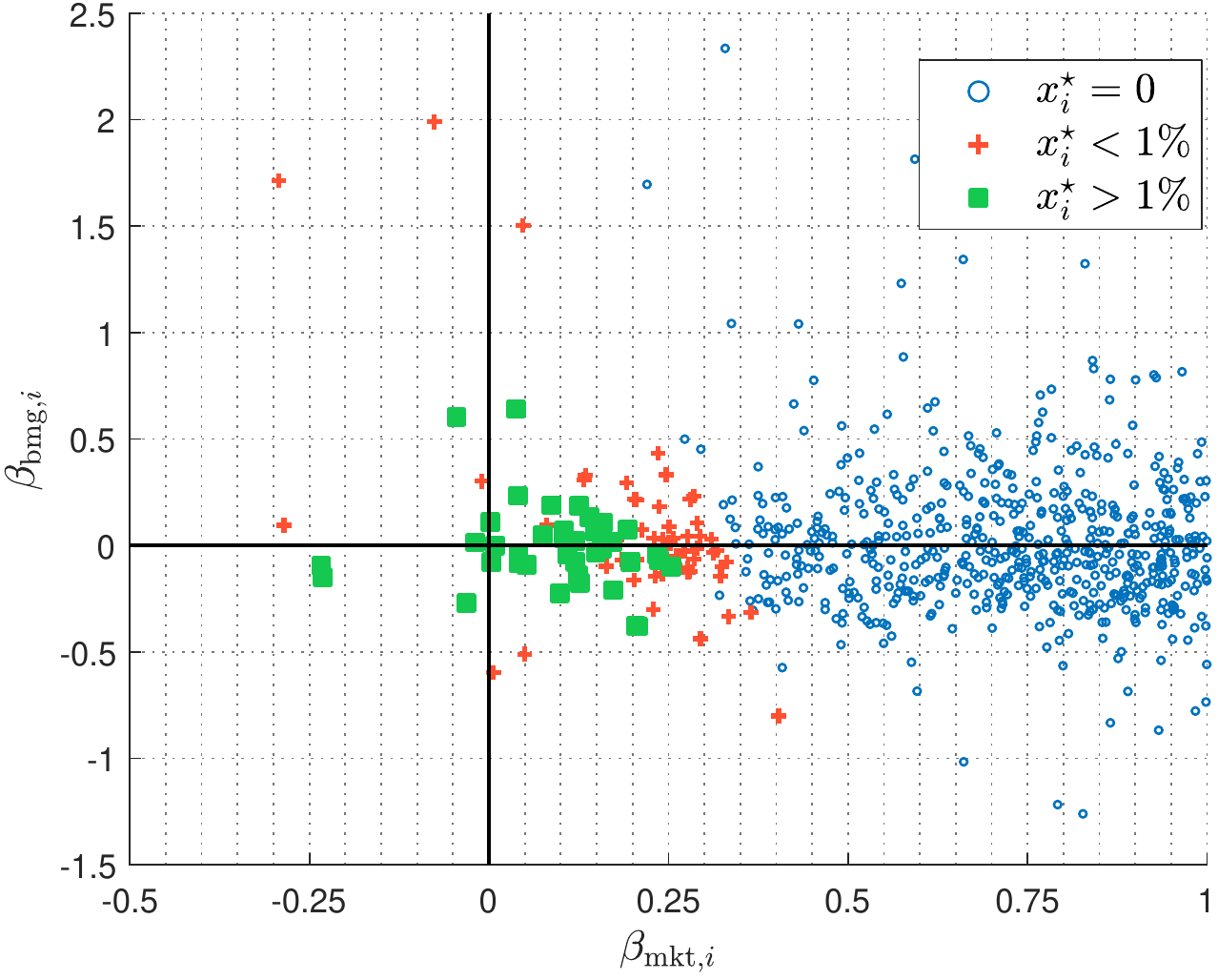}
\end{table}

\begin{table}[p]
\centering
\caption{Weights of the constrained MV portfolio ($\beta_{\mathrm{bmg}}^{+} = -0.25$)}
\label{fig:jpm_carbon7}
\smallskip

\includegraphics[width = \figurewidth, height = \figureheight]{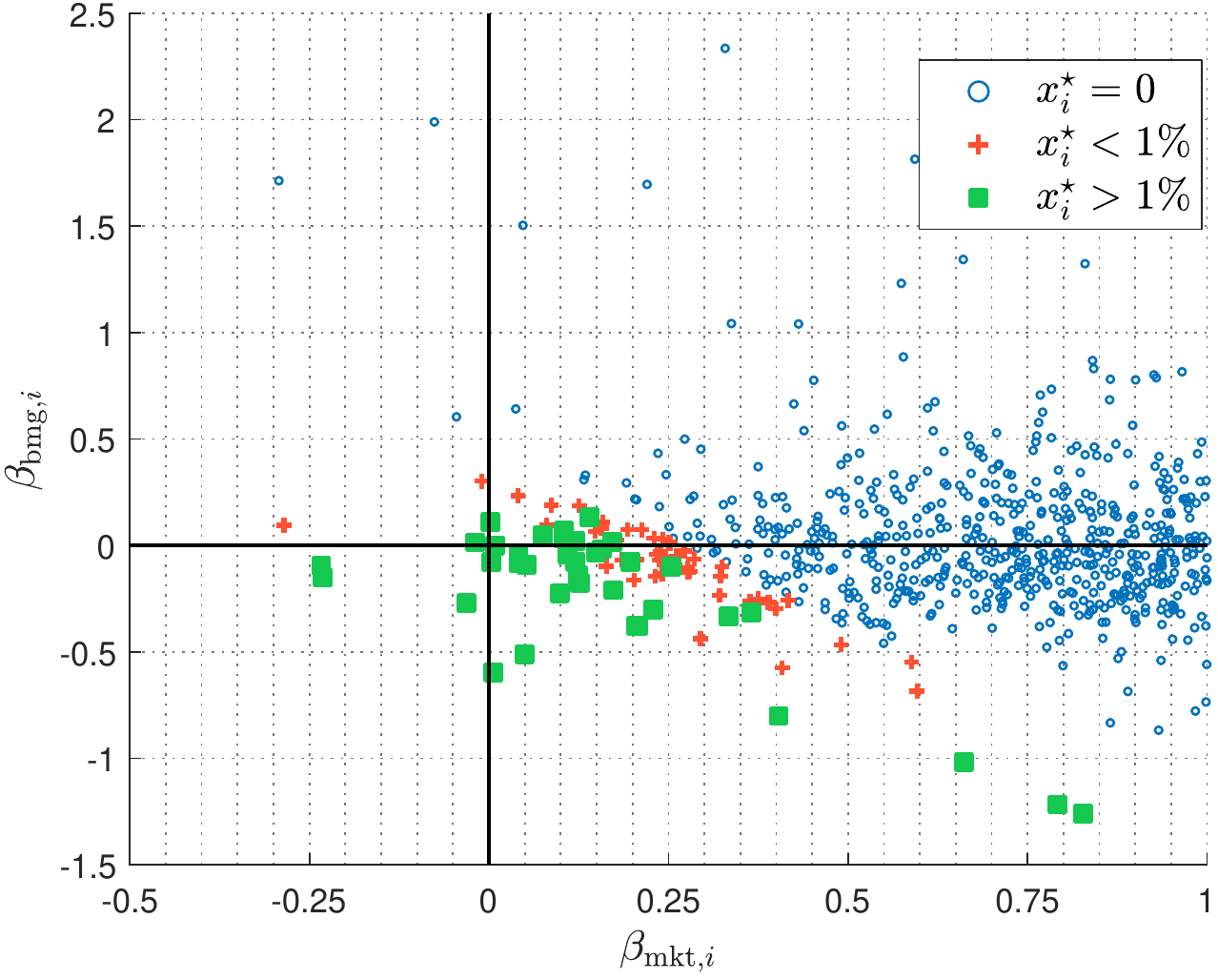}
\end{table}

\paragraph{Considering relative carbon risk}

In order to circumvent the previous drawback, we can directly add a
constraint in the optimization program:%
\begin{equation}
\beta _{\mathrm{bmg}}\left( x\right) =\sum_{i=1}^{n}x_{i}\times \beta_{\mathrm{bmg},i}
\leq \beta _{\mathrm{bmg}}^{+}
\end{equation}%
where $\beta _{\mathrm{bmg}}\left( x\right) $ is the carbon beta of
portfolio $x$ and $\beta _{\mathrm{bmg}}^{+}$ is the maximum
tolerance of the investor with respect to the relative carbon risk.
We consider the previous example. If we would like to impose a
carbon sensitivity of lower than $-0.25$, we obtain results given in
Exhibit \ref{fig:jpm_carbon7}. The comparison with Exhibit
\ref{fig:jpm_carbon6} shows that the MV portfolio tends to select
stocks with both a low market sensitivity and a negative carbon
beta. Moreover, large weights are associated with large negative
values of $\beta _{\mathrm{bmg},i}$ on average.

\paragraph{Managing both market and fundamental risk measures}

The previous method is not the standard approach when managing carbon risk
in investment portfolios. Indeed, the asset management industry generally
considers constraints on carbon intensity measures. Following Andersson et
al. (2016), we can impose individual constraints on the different stocks:
\begin{equation}
x_i = 0 \quad \text{if}\quad \mathcal{CI}_{i}\leq \mathcal{CI}^{+}
\end{equation}%
or we can use a global constraint:%
\begin{equation}
\mathcal{WACI}\left( x\right) =\sum_{i=1}^{n}x_{i}\times \mathcal{CI}_{i}\leq
\mathcal{WACI}^{+}
\end{equation}%
where $\mathcal{CI}_{i}$ is the carbon intensity of stock $i$ and
$\mathcal{WACI}\left( x\right) $ is the weighted average carbon intensity of portfolio
$x$. $\mathcal{CI}^{+}$ and $\mathcal{WACI}^{+}$ are the individual and portfolio
thresholds that are accepted by investors.\smallskip

We may wonder whether managing the fundamental measure of carbon
risk is equivalent to managing the market measure of carbon
risk\footnote{In what follows, we also impose that $\mathcal{CI}^{+}
= 4\,000$.}. A preliminary answer has been provided previously since
we have found that the correlation between $\beta _{\mathrm{bmg},i}$
and $\mathcal{CI}_{i}$ is less than $30\%$ on average. In Exhibit
\ref{tab:jpm_carbon8a}, we compute the minimum variance portfolio by
considering several threshold values of $\beta _{\mathrm{bmg}}^{+}$.
We notice that using a lower value of $\beta _{\mathrm{bmg}}^{+}$
reduces the value of $\mathcal{WACI}\left( x\right) $, but
$\mathcal{WACI}\left( x\right) $ remains very high because some
issuers have a low common carbon risk, but a high idiosyncratic
carbon risk. We have also reported the number of stocks
$\mathcal{N}\left( x\right) $ in the MV portfolio. As expected, it
decreases when we impose a stronger constraint. Exhibit
\ref{tab:jpm_carbon8b} is a variant of Exhibit
\ref{tab:jpm_carbon8a} by considering a constraint
$\mathcal{WACI}^{+}$ on the portfolio's carbon intensity instead of
a constraint $\beta _{\mathrm{bmg}}^{+}$ on the portfolio's carbon
beta. Here, the impact on the portfolio's carbon beta is low when we
strengthen the constraint. Indeed, the portfolio's carbon beta
$\beta _{\mathrm{bmg}}\left( x\right) $ is equal to 1.43\% when we
target a carbon intensity of 500, whereas it drops to 1.33\% when
the constraint on the carbon intensity is set to 50.\smallskip

\begin{table}[h]
\centering
\caption{Minimum variance portfolios with a relative carbon beta constraint}
\label{tab:jpm_carbon8a}
\smallskip

\begin{tabular}{rrcr}
\hline
$\beta_{\mathrm{bmg}}^{+}$ & $\beta_{\mathrm{bmg}}\left(x\right)$ &
$\mathcal{WACI}\left(x\right)$ & $\mathcal{N}\left(x\right)$ \\
\hline
         &   1.43\% & 538 & 105 \\
-10.00\% & -10.00\% & 501 & 100 \\
-20.00\% & -20.00\% & 422 &  89 \\
-40.00\% & -40.00\% & 289 &  70 \\
\hline
\end{tabular}
\end{table}

\begin{table}[h]
\centering
\caption{Minimum variance portfolios with a carbon intensity constraint}
\label{tab:jpm_carbon8b}
\smallskip

\begin{tabular}{rrcr}
\hline
$\mathcal{WACI}^{+}$ & $\mathcal{WACI}\left(x\right)$ &
$\beta_{\mathrm{bmg}}\left(x\right)$ & $\mathcal{N}\left(x\right)$ \\
\hline
500 & 500 & 1.43\% & 105 \\
250 & 250 & 1.37\% & 103 \\
100 & 100 & 1.36\% &  98 \\
 50 &  50 & 1.33\% &  82 \\
\hline
\end{tabular}
\end{table}

These results show that the two optimization problems give two different
solutions in terms of carbon risk. Therefore, it makes sense to combine the
approaches by imposing two constraints:
\begin{equation}
\left\{
\begin{array}{l}
\mathcal{WACI}\left( x\right) \leq \mathcal{WACI}^{+} \\
\beta _{\mathrm{bmg}}\left( x\right) \leq \beta _{\mathrm{bmg}}^{+}
\end{array}%
\right.
\end{equation}
Moreover, the threshold $\beta _{\mathrm{bmg}}^{+}$ allows us to
reduce the common carbon risk, but not the idiosyncratic carbon
risk. The WACI constraint circumvents this problem. Exhibit
\ref{tab:jpm_carbon8c} presents the results for several values of
$\mathcal{WACI}^{+}$ when $\beta _{\mathrm{bmg}}^{+}$ is equal to
$-20\%$. For instance, we notice that the WACI constraint is not
reached when $\mathcal{WACI}^{+}=500$ and $\beta
_{\mathrm{bmg}}^{+}=-20\%$. The last column of Exhibit
\ref{tab:jpm_carbon8c} corresponds to the portfolio's weight overlap
with respect to the optimized portfolio based on the WACI
constraint, meaning that we compare the portfolio optimized with the
BMG and WACI constraints to the portfolio optimized with the WACI
constraint. In this example, we notice that the weight overlap
$\mathcal{WO}\left( x\right) $ is equal to $75\%$ on average. This
means that $25\%$ of the minimum variance portfolio allocation is
changed when we add the market carbon constraint $\beta
_{\mathrm{bmg}}^{+}=-20\%$.

\begin{table}
\centering
\caption{Minimum variance portfolios with carbon beta and intensity constraints}
\label{tab:jpm_carbon8c}
\smallskip

\begin{tabular}{rrcrc}
\hline
$\mathcal{WACI}^{+}$ & $\mathcal{WACI}\left(x\right)$ &
$\beta_{\mathrm{bmg}}\left(x\right)$ & $\mathcal{N}\left(x\right)$ &
$\mathcal{WO}\left(x\right)$ \\
\hline
500 & 430 & -20.00\% & 111 & 74.65\% \\
250 & 250 & -20.00\% &  86 & 75.26\% \\
100 & 100 & -20.00\% &  79 & 74.87\% \\
 50 &  50 & -20.00\% &  74 & 74.99\% \\
\hline
\end{tabular}
\end{table}

\section{CONCLUSION}

This paper considers the seminal approach of G\"{o}rgen et al.
(2019) to measuring carbon risk. While many asset managers and
owners use carbon intensity, we focus on the carbon beta which is
priced in by the market. The carbon beta is estimated using a
two-step approach. First, we build a brown-minus-green risk factor.
Second, we perform a Kalman filtering in order to obtain the
time-varying carbon beta. By considering this dynamic framework, we
highlight several stylized facts. We show that this market measure
is very different from a traditional fundamental measure of the
carbon risk. The main reason is that carbon intensity is not the
only dimension that is priced in by the market.\smallskip

Another important result is the difference between relative and
absolute carbon risk. Investors that are sensitive to relative
carbon risk prefer stocks with a negative carbon beta over stocks
with a positive carbon beta, whereas investors that are sensitive to
absolute carbon risk prefer stocks with a carbon beta close to zero.
Managing relative carbon risk implies having a negative exposure to
the carbon risk factor, whereas managing absolute carbon risk
implies having zero exposure to the carbon risk factor. The first
case is an active management bet since the performance may be
negative if brown stocks outperform green stocks. Nevertheless, this
approach reduces exposure to firms that face a threat of
environmental regulation (Maxwell et al., 2000). The second case is
an immunization investment strategy against carbon risk. However,
this hedging strategy is not widely implemented by institutional and
passive investors because of their moral values and convictions, and
they generally prefer to implement relative carbon risk strategies.\smallskip

Introducing carbon risk into a minimum variance portfolio is a hot
topic among asset managers and owners. Indeed, the goal of a minimum
variance portfolio is to build a low volatility strategy on the
equity market. This is achieved by considering a strong risk
management approach on several dimensions. Originally, the strategy
only focused on the portfolio's volatility. Since the 2008 Global
Financial Crisis, it has included other risk dimensions that can
burst the equity market such as credit risk and valuation risk.
Climate risk has become another important dimension, especially
because minimum variance strategies are massively implemented by ESG
institutional investors. In this context, the question of carbon
metrics is important. In this paper, we show that managing the
carbon intensity of minimum variance portfolios has little impact on
their carbon beta. The opposite is not true, but the effect of
managing the carbon beta on carbon intensity is limited. This is why
we propose combining the market and fundamental approaches to carbon
risk. Another issue concerns the choice of the market carbon risk
measure. We show that the optimization program of a minimum variance
portfolio naturally considers absolute carbon risk. However,
relative carbon risk can also be an option if the investor's goal is
not to hedge the carbon risk, but to be a green investor.

\section{REFERENCES}

\noindent
Andersson, M., P. Bolton, and F. Samama.
2016.
\textquotedblleft Hedging Climate Risk.\textquotedblright\
\textit{Financial Analysts Journal} 72 (3): 13-32.
\bigskip

\noindent
Bouchet, V., and T. Le Guenedal.
2020.
\textquotedblleft Credit Risk Sensitivity to Carbon Price.\textquotedblright\
\textit{SSRN} 3574486.
\bigskip

\noindent
Carney, M.
2015.
\textquotedblleft Breaking the Tragedy of the Horizon -- Climate Change and Financial Stability.\textquotedblright\
\textit{Bank of England Speeches on climate Change} 29 September 2015.
\bigskip

\noindent
Carney, M.
2019.
\textquotedblleft Remarks given during the UN Secretary General's Climate Action Summit 2019.\textquotedblright\
\textit{Bank of England Speeches on climate Change} 24 September 2019.
\bigskip

\noindent
Clarke, R.G., H. de Silva, and S. Thorley.
2011.
\textquotedblleft Minimum Variance Portfolio Composition.\textquotedblright\
\textit{Journal of Portfolio Management} 37 (2): 31-45.
\bigskip

\noindent
Fabozzi, F. J., and J. C. Francis.
1978.
\textquotedblleft Beta as a Random Coefficient.\textquotedblright\
\textit{Journal of Financial and Quantitative Analysis} 13 (1):101-116.
\bigskip

\noindent
Fama, E. F., and K. R. French.
1992.
\textquotedblleft The Cross-Section of Expected Stock Returns.\textquotedblright\
\textit{The Journal of Finance} 47 (2): 427-465.
\bigskip

\noindent
G\"orgen, M., A. Jacob, M. Nerlinger, R. Riordan, M. Rohleder, and M. Wilkens.
2019.
\textquotedblleft Carbon Risk.\textquotedblright\
\textit{SSRN} 2930897.
\bigskip

\noindent
Maxwell, J. W., T.P. Lyon, and S. C. Hackett.
2000.
\textquotedblleft  Self-regulation and Social Welfare: The Political Economy of
Corporate Environmentalism.\textquotedblright\
\textit{Journal of Law and Economics}, 43 (2): 583-618.
\bigskip

\noindent
MSCI.
2020.
\textquotedblleft MSCI ESG Ratings Methodology.\textquotedblright\
\textit{MSCI ESG Research} April 2020.
\bigskip

\noindent
Roncalli, T., T. Le Guenedal, F. Lepetit, T. Roncalli, and T. Sekine
2020.
\textquotedblleft Measuring and Managing Carbon Risk in Investment Portfolios.\textquotedblright\
\textit{SSRN} 3681266.\bigskip

\noindent
Scherer, B.
2011.
\textquotedblleft A Note on the Returns from Minimum Variance Investing.\textquotedblright\
\textit{Journal of Empirical Finance} 18 (4):652-660.

\end{document}